\documentclass[aps,prb,a4paper,10pt,twocolumn,floatfix,superscriptaddress,amsmath,longbibliography,preprintnumbers,citeautoscript]{revtex4-1}
\usepackage[T1]{fontenc}
\usepackage{dcolumn,graphicx,color,booktabs,microtype,afterpage}

\graphicspath{{./}{figure/}}
\usepackage[charter,greekuppercase=italicized]{mathdesign}
\renewcommand{\tablename}{Table}
\makeatletter\renewcommand{\fnum@figure}[1]{\figurename~\thefigure.~}\makeatother
\makeatletter\renewcommand{\fnum@table}[1]{\tablename~\thetable.~}\makeatother

\newcount\hh \newcount\mm
\hh=\time \divide\hh by 60
\mm=\hh \multiply\mm by 60 \mm=-\mm
\advance\mm by \time
\def\now{\number\hh:\ifnum\mm<10{}0\fi\number\mm}

\usepackage[colorlinks,plainpages=false,linkcolor=blue,urlcolor=blue,citecolor=blue,pdfpagemode=UseNone,pdfstartview=FitBH]{hyperref}

\usepackage{nicefrac}
\newcommand{\BACSO}{Ba\-Cu$_2$\-Si$_2$\-O$_7$}

\newcommand{\BASGE}{Ba\-Cu$_2$\-Si\-Ge\-O$_7$}
\newcommand{\BASGX}{Ba\-Cu$_2$\-(Si$_{1-x}$\-Ge$_x$)$_2$\-O$_7$}

\begin{document}

\makeatletter\renewcommand{\ps@plain}{%
\def\@evenhead{\hfill\itshape\rightmark}%
\def\@oddhead{\itshape\leftmark\hfill}%
\renewcommand{\@evenfoot}{\hfill\small{--~\thepage~--}\hfill}%
\renewcommand{\@oddfoot}{\hfill\small{--~\thepage~--}\hfill}%
}\makeatother\pagestyle{plain}


\title{From order to randomness: onset and evolution of the random-singlet state\\ in bond-disordered 
\texorpdfstring{BaCu$_2$(Si$_{1-x}$Ge$_x$)$_2$O$_7$}{BaCu2(Si(1-x)Gex)2O7} spin-chain compounds}

\author{T.\,Shiroka}\email[Corresponding author: \vspace{8pt}]{tshiroka@phys.ethz.ch}
\affiliation{Laboratorium f\"ur Festk\"orperphysik, ETH H\"onggerberg, CH-8093 Z\"urich, Switzerland}
\affiliation{Paul Scherrer Institut, CH-5232 Villigen PSI, Switzerland}

\author{F.\,Eggenschwiler}
\affiliation{Laboratorium f\"ur Festk\"orperphysik, ETH H\"onggerberg, CH-8093 Z\"urich, Switzerland}

\author{H.-R.\,Ott}
\affiliation{Laboratorium f\"ur Festk\"orperphysik, ETH H\"onggerberg, CH-8093 Z\"urich, Switzerland}
\affiliation{Paul Scherrer Institut, CH-5232 Villigen PSI, Switzerland}

\author{J.\,Mesot}
\affiliation{Laboratorium f\"ur Festk\"orperphysik, ETH H\"onggerberg, CH-8093 Z\"urich, Switzerland}
\affiliation{Paul Scherrer Institut, CH-5232 Villigen PSI, Switzerland}

\begin{abstract}
\noindent 
Heisenberg-type spin-chain materials have been extensively studied over the years, 
yet not much is known about their behavior in the presence of disorder. Starting 
from BaCu$_2$Si$_2$O$_7$, a typical spin-$\nicefrac{1}{2}$ chain system, we 
investigate a series of compounds with different degrees of bond disorder, where 
the systematic replacement of Si with Ge results in a re-modulation of the Cu$^{2+}$ 
exchange interactions.
By combining magnetometry measurements with nuclear magnetic resonance studies 
we follow the evolution of the disorder-related properties from the well-ordered 
BaCu$_2$Si$_2$O$_7$ to the maximally disordered BaCu$_2$SiGeO$_7$. 
Our data indicate that already a weak degree of disorder of only 5\% Ge, apart 
from reducing the 3D magnetic ordering temperature $T_\mathrm{N}$ quite 
effectively, induces a qualitatively different state in the paramagnetic regime. 
At maximum disorder our data indicate that this state may be identified with 
the theoretically predicted random singlet (RS) state. 
With decreasing disorder the extension of the RS regime at temperatures 
above $T_\mathrm{N}$ is reduced, yet its influence is clearly manifest, 
particularly in the features of NMR relaxation data.  
\end{abstract}



\maketitle\enlargethispage{3pt}

\vspace{-5pt}\section{Introduction}\enlargethispage{8pt}
The study of electronic properties of physical systems in the presence 
of disorder spans many decades,\cite{Abrahams2010} starting with 
the \emph{strong} (Anderson) \emph{localization} studies in the 1950s\cite{Anderson1958} 
up to the present-day investigations of quantum confinement in nanostructures.\cite{Oka2014}
The breadth of phenomena taking place in disordered systems, such as 
quantum percolation,\cite{Schubert2009} ballistic transport,\cite{Stevens1987} 
quantum glassiness,\cite{Chamon2005} or many-body localization\cite{Alet2018} 
have been studied primarily theoretically, e.g., as a function of dimensionality, 
nature of disorder, degree of interaction, etc. Of particular interest is the physics 
occurring in low-dimensional quantum magnets under a varying degree of disorder. 
At very low temperatures and high magnetic fields, close to a quantum phase 
transition, disorder suppresses the global phase coherence and induces novel 
quantum critical behavior.\cite{Zheludev2013,Zapf2014}
But even under less extreme conditions, the disorder-induced breaking of 
translational invariance promotes random couplings between individual spins 
and leads to a so-called random-singlet (RS) state,\cite{Dasgupta1980,Fisher1994,Motrunich2000} 
a regime where spins couple across arbitrary distances to form 
weakly-bound singlets, which dominate the magnetic features and the related dynamics.  
What exactly happens when a regular spin-chain is exposed to an 
increasing degree of disorder is not well known. 
Until recently, progress has been slow as far as numerical simulations\cite{Shu2018} and, 
especially, experimental investigations\cite{Hammerath2011,Hlubek2010,Utz2017}  
of \emph{disordered} low-dimensional systems are concerned. 
The main reasons include computational difficulties due to the large size 
of realistic disordered systems and, regarding experiments, the scarcity 
of suitable systems in which disorder can be easily tuned over a broad range 
without changing the structural character of the material.

It is known for some time that the series of \BASGX\ compounds represents 
one of the best physical realizations of an $S = 1/2$ Heisenberg-type 
spin-chain system (see, e.g., Table 1 in Ref.~\onlinecite{Broholm2002}), 
where bond disorder can be introduced in a controlled way. For $x = 0$, 
the compound crystallizes in the orthorhombic space group \textit{Pnma} 
($D^{16}_{2h}$) with lattice constants $a = 6.862$\,\AA, $b = 13.178$\,\AA, 
and $c = 6.897$\,\AA.\cite{Yamada2001} 
Replacing Si by Ge results in isostructural compounds across the entire series, 
but introduces a variation in the O-Cu-O bond angle from 124$^{\circ}$ in 
the Si case ($x = 0$) to 135$^{\circ}$ in the Ge case ($x = 1$).\cite{Yamada2001} 
As a consequence, the exchange-coupling constant almost doubles, from 
$J = 24.1$\,meV to 46.5\,meV, when $x$ changes from 0 to 1.\cite{Tsukada1999,Kenzelmann2001} 
In this way the randomization, in the form of varying couplings, enters 
the magnetically relevant Cu-O chains. The parent compound \BACSO\ orders 
antiferromagnetically at $T_{\mathrm{N}} = 9.2$\,K, i.e., at a much lower 
temperature than 280\,K, the equivalent of the exchange energy $J = 24.1$\,meV. 
Since also the $x >0$ members have similar (or lower) $T_{\mathrm{N}}$ 
values, this qualifies the \BASGX\ series as one of the best 1D systems 
for studying bond-disorder effects.

\begin{figure}[t]
\includegraphics[width=0.8\columnwidth]{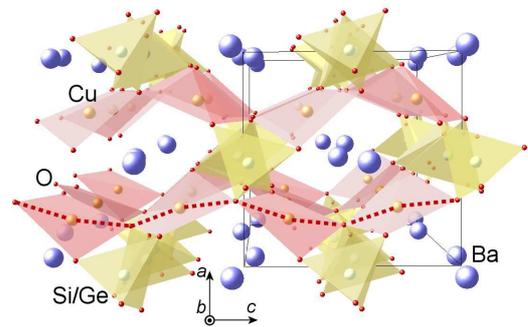} 
\caption{Schematic crystal structure of BaCu$_2$(Si$_{1-x}$Ge$_x$)$_2$O$_7$. 
The corner-sharing CuO$_2$ chains (dotted line) run along the $c$ axis. The 
random presence of tetrahedrally-coordinated Si${}^{4+}$ and Ge${}^{4+}$ 
ions modulates $J$, the interaction coupling strength between Cu${}^{2+}$ ions.}
\label{Fig:struct}
\end{figure}

In previous studies, concerning the effects of randomness on the compound's 
magnetic properties, we focused on the extreme cases of complete order 
or disorder, i.e., $x = 0$ and $x = 0.5$.\cite{Shiroka2011,Shiroka2013}  
The main result of bulk magnetization measurements and of low-temperature 
${}^{29}$Si NMR studies, supplemented by quantum Monte Carlo calculations, 
was the experimental verification of the theoretically predicted \emph{random-singlet} 
state\cite{Fisher1994,Dasgupta1980} in the $x = 0.5$ case. 
Important ingredients of this identification are the experimental determination 
of the probability distribution of NMR spin-lattice relaxation times $T_1$ and the 
rather unexpected (and so far unexplained) temperature dependence of $T_1^{-1}$ 
for $x = 0.5$, characteristically different from that of the $x = 0$ case. 
The current work aims to investigate the gradual development of the earlier 
established features, characteristic of the formation of the random-singlet 
state, as $x$ is varied between 0 and 0.5. To this end we employ the same experimental 
techniques. In particular, we use the distinct variation of $T_1^{-1}(T)$ in the 
$0 < x < 0.5$ range to identify the unexpectedly rapid formation of the random-singlet 
state in a limited temperature interval, which grows with increasing $x$.

The paper is organized as follows: in Sec.~\ref{sec:exp_details} we present 
the experimental details of sample growth and characterization, as well as 
those related to the measurement techniques. 
The magnetometry and NMR data are presented and analyzed in 
Sec.~\ref{sec:analysis}. In Sec.~\ref{sec:discussion} we discuss our 
main findings and compare them with those of recent muon-spin rotation 
studies.\cite{Thede2014} Finally, in Sec.~\ref{sec:conclusion}, we summarize 
our key results.

\vspace{-5pt}\section{Experimental details\label{sec:exp_details}}
Single crystals of \BASGX\ with $x=0$, 0.05, 0.15, 0.30, 0.50, and 0.95 
were grown by the floating-zone technique, using stoichiometric amounts 
of reactants. Subsequently, laboratory-based x-ray diffraction (Bruker AXS 
D8 Discover) was used to characterize the samples, as well as to identify 
their orientation. Our data confirm that the considered barium copper compounds 
share the same orthorhombic \textit{Pnma} space group,\cite{Yamada2001} 
with only a tiny increase (2\%) in lattice parameters from the silicate 
($x=0$) to the germanate ($x=1$) case.
For all the samples, detailed structural analyses showed the excellent 
agreement between the nominal and the real Ge concentration $x$, as 
well as its homogeneous distribution within each sample.

The magnetic susceptibility was measured via standard dc magnetometry 
(SQUID, Quantum Design) at different values of the applied magnetic 
field (0.01, 0.1, and 1 T) and at temperatures between 4 and 300\,K. 
Since the $a$ and $c$ lattice parameters are very similar, only the $b$ 
axis can be easily identified. Therefore, for both the magnetometry 
and the NMR measurements the magnetic field was aligned along the 
crystalline $b$ axis.

The $^{29}$Si nuclear magnetic resonance line shapes and spin-lattice 
relaxation times $T_1$ were measured using standard pulsed techniques 
in an external field of $\mu_0H = 7.07$\,T. The NMR measurements were 
restricted to samples with $x \leq 0.5$, since at higher Ge content the  $^{29}$Si 
NMR signal progressively weakens. The field value was calibrated by means 
of an aluminum sample. The well established reference frequency of 
${}^{27}$Al nuclei was used to determine the absolute position of the 
$^{29}$Si lines (typically at 59.715\,MHz), as well as to track their shifts 
with varying temperature. 
A typical pulse duration of $\sim 5$\,$\mu$s  was sufficient to irradiate 
the entire resonance, except at low temperatures in the $x = 0.5$ case, 
where disorder-broadened NMR lines required frequency scans. 
For a comparison with previous results, the same NMR measurements 
were carried out both on aligned single crystals and on powder samples 
with the same chemical composition. The relatively small differences which 
were detected, can be ascribed to the presence of crystalline anisotropies.

\vspace{-5pt}\section{Results and data analysis\label{sec:analysis}}
\subsection{Magnetization measurements}
\begin{figure}[t]
\includegraphics[width=0.95\columnwidth]{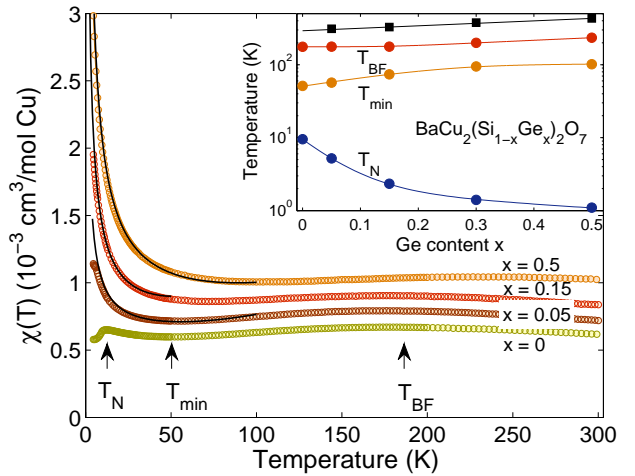} 
\caption{Static magnetic susceptibility per mole Cu vs.\ temperature at 
$\mu_0H = 100$ mT for a selection of \BASGX\ samples with different 
Ge-substitution values $x$. Full lines are fits by means of Eq.~(\ref{eq:cutoff}). 
To improve the visibility the curves are vertically shifted by 0.2 units. 
Key features, such as the N\'eel temperature, the minimum, and the 
Bonner-Fisher peak are marked by arrows and their variation vs.\ Ge 
content is shown in the inset together with that of the fitted  $\Omega/k_\mathrm{B}$ 
values (black squares). Since the $x=0.3$ sample shows very similar 
features to $x =0.5$, here and elsewhere we show only the metadata 
for the $x=0.3$ case.}
\label{Fig:chi_T_x}
\end{figure}

The macroscopic magnetic response of samples with different Ge 
substitution was probed by measurements of the magnetization $M$, 
both as a function of temperature and field. In the first case, clear and 
distinct changes were noted upon Ge substitution $x$, as illustrated in 
Fig.~\ref{Fig:chi_T_x}, where data of the magnetic susceptibility 
$\chi_{x}(T)$ are displayed. 
All curves exhibit a broad (Bonner-Fisher-type) maximum,\cite{Bonner1964} 
a typical feature of one-dimensional spin systems. Its position, located 
at 180\,K for the $x=0$ case, shifts towards higher temperatures as 
the Ge content is enhanced. Since the susceptibility maximum is 
expected to occur at $k_\mathrm{B}T_\mathrm{max} = 0.641 J$ 
and to reach a value of $\chi_\mathrm{max} \simeq 0.147 N g^2 \mu^2_\mathrm{B} / J$ 
[see Eq.~(30a) in Ref.~\onlinecite{Johnston2000}], a gradual increase 
of $T_\mathrm{max}$ (i.e., $T_{\mathrm{BF}}$ in Fig.~\ref{Fig:chi_T_x}) 
and the concomitant decrease of the corresponding susceptibility 
maximum clearly reflect an increase in average $J$ as the Ge content 
$x$ is enhanced. 
By using $T_{\mathrm{max}} = 0.641\,J/ k_{\mathrm{B}}$, strictly valid 
only in the uniform-$J$ case, we obtain $J/k_{\mathrm{B}} = 280.8$\,K, 
in excellent agreement with previously derived values from magnetometry\cite{Yamada2001} 
or neutron diffraction studies ($J = 24.1$\,meV, equivalent to 280\,K).\cite{Tsukada1999}
For $x>0$, the same approach can still provide an average $J$ value, yet 
it cannot fully capture the true nature of the disorder-induced effects (see below).

The sharp maximum in $\chi(T)$ at low temperatures indicates the transition to 
a 3D antiferromagnetic state,\footnote{While at low $x$ values $T_{\mathrm{N}}$ 
is clearly visible from the $\chi(T)$ data, at higher $x$ values, due to the 
rapid divergence of $\chi(T)$, it appears only as a small change in slope.} 
with the value of $T_{\mathrm{N}}$ reflecting the intensity of the residual 
intrachain interactions. The position of the $T_{\mathrm{N}}$ peak shifts 
towards lower temperatures as the Ge substitution is enhanced (see also 
Ref.~\onlinecite{Yamada2001}). The temperature dependence of 
$T_{\mathrm{N}}$,  $T_{\mathrm{BF}}$, as well as that of an intermediate 
minimum, are shown in the inset of Fig.~\ref{Fig:chi_T_x}.

Measurements of the magnetization $M(H)$ as a function of field, from 
$\mu_0H = 0$ up to 5\,T, were also performed at selected temperatures. 
A change in applied field by two orders of magnitude did not show essential 
variations in susceptibility, except for a better signal-to-noise ratio at 
higher fields. The $M(H)$ data (not shown here) reveal a linear variation 
of $M$ vs.\ $H$, with an increasing slope as the temperature decreases.
The observed linearity indicates the low effective magnetic moment of 
Cu$^{2+}$ ions, a typical feature of the quantum magnetism of spin chains.
 
Fitting the susceptibility data $\chi(T)$ is not straightforward, since analytical 
formulas exist only in the limiting cases of a uniform $J$ ($x = 0$), or of an 
alternating-exchange $J_1$-$J_2$ ($x = 0.5$) Heisenberg spin-chain model.
In earlier work, magnetometry data of qua\-si\--1D spin chain systems 
were analyzed by using the expression\cite{Yamada2001,Johnston2000}
\begin{equation}
\chi(T) = \chi_0 + \chi_{\mathrm{CW}}(T) + \chi_{\mathrm{spin}}(T).
\label{eq:fit_magn}
\end{equation}
Here $\chi_0$ is a temperature-independent term due to core diamagnetism 
and Van Vleck paramagnetism, $\chi_{\mathrm{CW}} = C/(T-\Theta)$ is 
a Curie-Weiss contribution, and $\chi_{\mathrm{spin}}(T)$ represents the 
susceptibility of the 1D spin system. The Curie-type term was introduced 
to account for the low-temperature upturn, typically being attributed to 
impurities, defects, or intrinsic weak ferromagnetism.

Although formal fits using Eq.~(\ref{eq:fit_magn}) follow fairly well the 
experimental data, this approach has some serious drawbacks.
Firstly, it fails to capture the physical meaning of the low-$T$ divergence, 
which in our case is \emph{not} due to impurities, but rather reflects the 
low-energy states associated with the weakly-bound singlets.
In the theoretically predicted random-singlet state, their magnetic 
response is modeled by:\cite{Fisher1994}
\begin{equation}
\chi(T) \sim T^{-1} \ln^{-2}(\Omega/T),
\label{eq:cutoff}
\end{equation}
with $\Omega$ a cutoff energy. In previous work, we showed that for 
$x=0.5$, Eq.~(\ref{eq:cutoff}) describes very well the upturn in $\chi(T)$, 
with a reasonable value of the fit parameter $\Omega$.\cite{Shiroka2011} 
Secondly,  Eq.~(\ref{eq:fit_magn}) provides only an \emph{average}, 
$x$-dependent exchange coupling value $J$ which, in Ref.~\onlinecite{Shiroka2011}, 
was clearly shown to be insufficient to capture the complexity of spin 
chains with bond randomness.

As demonstrated in Fig.~\ref{Fig:chi_T_x}, by using Eq.~(\ref{eq:cutoff}) 
we obtain reasonably good fits for all the $x >0$ susceptibility curves. 
Obviously, to capture the influence of randomness, we had to restrict the fit 
range from ca.\ $1.5 T_\mathrm{N}$ to $0.5 T_\mathrm{BF}$. This choice 
is justified because it avoids the complications of the 3D ordering effects 
at very low temperatures, while still exploring the quantum effects of the 
random-singlet state, which lose dominance at higher temperatures. 
The inset of Fig.~\ref{Fig:chi_T_x} shows the variation of $\Omega$ 
with $x$ (black squares).

\subsection{Magnetic resonance measurements}
\begin{figure}[t]
\includegraphics[width=0.75\columnwidth]{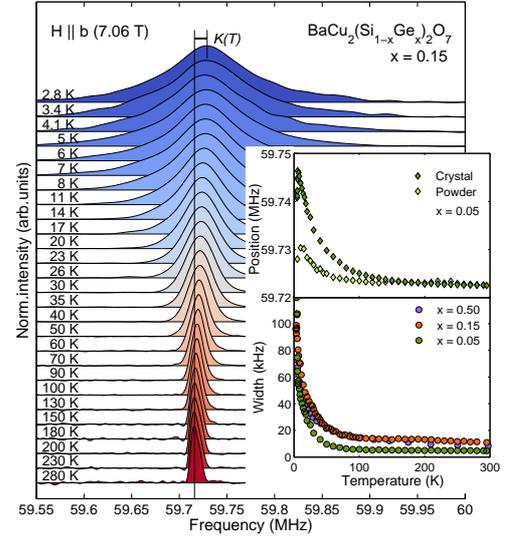} 
\caption{\label{Fig:lines_15}Evolution of $^{29}$Si NMR lines with temperature 
in \BASGX\ for an $x = 0.15$, single-crystal sample with the $b$-axis aligned 
along the field. The upper inset shows typical lineshifts for the $x =0.05$ case, 
where both powder- and single-crystal data exhibit a cusp at $T_\mathrm{N}$ 
(other samples show similar features). 
The lower inset highlights the low-temperature increase of linewidth for 
different $x$ values. For $T > 100$\,K, the wider lines in the $x = 0.15$ and 
0.50 case indicate a  disorder-induced broadening.}
\end{figure}
To gain further insight into the low-temperature spin dynamics and the 
local magnetic susceptibility we performed detailed NMR measurements. 
Since the lineshift $K$ reflects the intrinsic spin susceptibility at zero 
frequency $\chi(q = 0, \omega = 0)$, while the spin-lattice relaxation 
rate tracks the (low-energy) spin fluctuations, NMR is among the best 
techniques for studying the static and dynamic properties of low-dimensional 
spin systems (see, e.g., Ref.~\onlinecite{Horvatic2002}). \\
%

\paragraph{NMR linewidths and shifts.}
The evolution of the $^{31}$P NMR lineshapes with temperature for an 
$x = 0.15$ single crystal is shown in Fig.~\ref{Fig:lines_15}. The linewidth 
is approximately constant at high temperatures, but exhibits a divergent 
increase starting at an onset of $\sim 80$\,K. Similar results were obtained 
also for other concentrations $x$, but the onset temperature increases as 
the Ge substitution level $x$ is enhanced. 
An NMR linewidth, rapidly varying with decreasing temperature, directly 
reflects the growing distribution of the local magnetic susceptibilities and 
is in good agreement with theoretical predictions for random Heisenberg 
chains.\cite{Fisher1994} In general, an increase in Ge content causes a 
growth of disorder, reflected in enhanced linewidths already at room 
temperature. This is clearly shown in the lower inset of Fig.~\ref{Fig:lines_15} 
where, at $T > 100$\,K (dominated by classical disorder effects), the FWHM 
linewidths in the $x = 0.15$ case are a factor of two larger than for $x = 0.05$. 
A further increase in disorder (for $x = 0.50$) does not imply an additional 
increase in width at $T > 100$\,K, but rather a FWHM saturation. This is 
to be contrasted with the faster decrease of the stretching coefficient 
$\beta$ with decreasing temperature at higher $x$ values (see below). 

While linewidths exhibit a significant increase at low $T$, lineshifts $K(T)$ 
are generally quite small. Even in case of single-crystalline samples, the 
total $^{29}$Si NMR lineshifts are less than 0.03\%, not surprising 
considering the 1D nature of \BASGX. 
At the same time, typical of 1D materials, the wide distribution of local 
Knight shifts of magnetic origin implies a strong increase of linewidths at 
low temperature.
As shown in Fig.~\ref{Fig:lines_15}, $K(T)$ represents only a tiny fraction 
of the linewidth, thus making it difficult to precisely evaluate its value. 
Lineshape measurements on powder samples show a similar behavior 
(upper inset in Fig.~\ref{Fig:lines_15}),  with small differences most 
likely ascribed to the presence of crystalline anisotropies.
For all $x > 0$ cases, $K(T)$ shows first a progressive increase with 
decreasing temperature, then it reaches a sharp cusp-like maximum at 
the respective ordering temperature $T_\mathrm{N}$. These results 
confirm (at a microscopic level) the magnetization data shown in 
Fig.~\ref{Fig:chi_T_x} and are consistent with predictions for spin-chains 
in local transverse staggered fields, as reported in previous work on 
single crystals of BaCu$_2$Si$_2$O$_7$ ($x = 0$).\cite{Casola2012} 
It is worthwhile noting that, unlike the $x = 0$ case, where below 
$T_\mathrm{N}$ the two components of the NMR line can easily be 
resolved (see Fig.~2 in Ref.~\onlinecite{Casola2012}), for $x > 0$, due 
to disorder-induced broadening, this is not possible. In this case, we have 
direct access only to the total shift, i.e., to the average shift of the two 
components, whose position is relatively insensitive to disorder. Indeed, the 
cusp-like maximum in $K(T)$ observed for $x > 0$ reproduces perfectly the 
\emph{average-shift} results reported for the $x = 0$ case.\cite{Casola2012} \\

\begin{figure}[t]
\includegraphics[width=0.8\columnwidth]{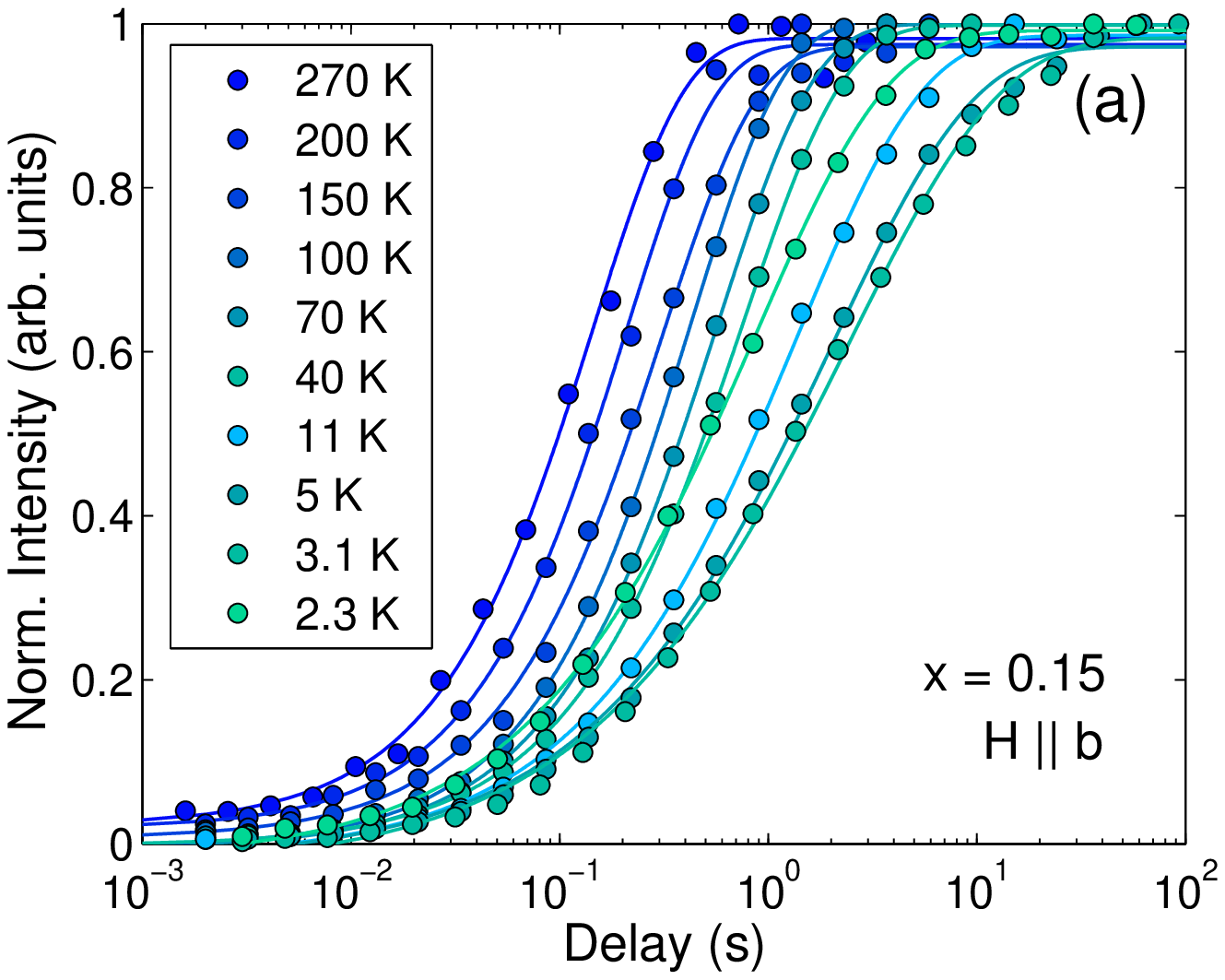} 
\includegraphics[width=0.8\columnwidth]{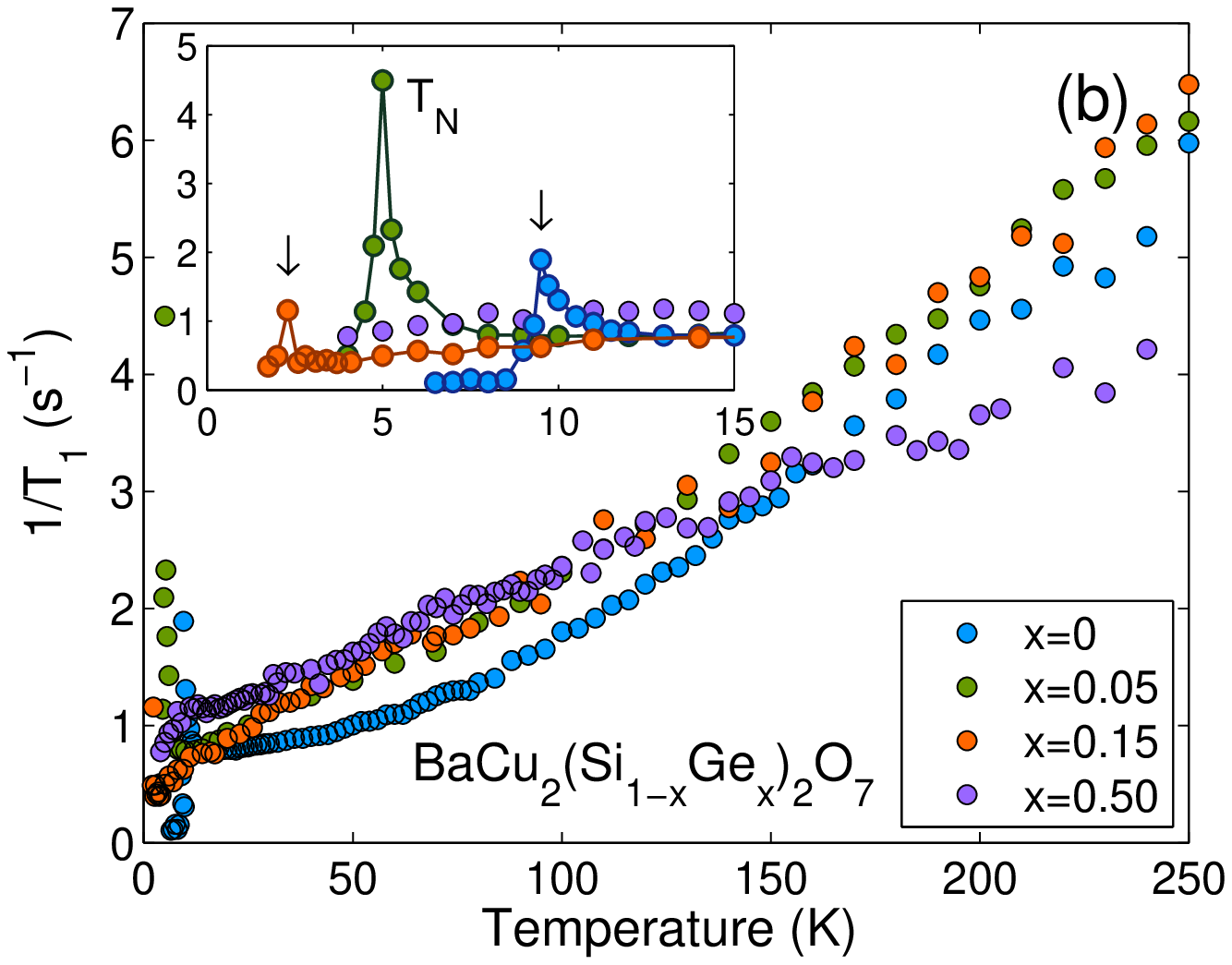} 
\includegraphics[width=0.8\columnwidth]{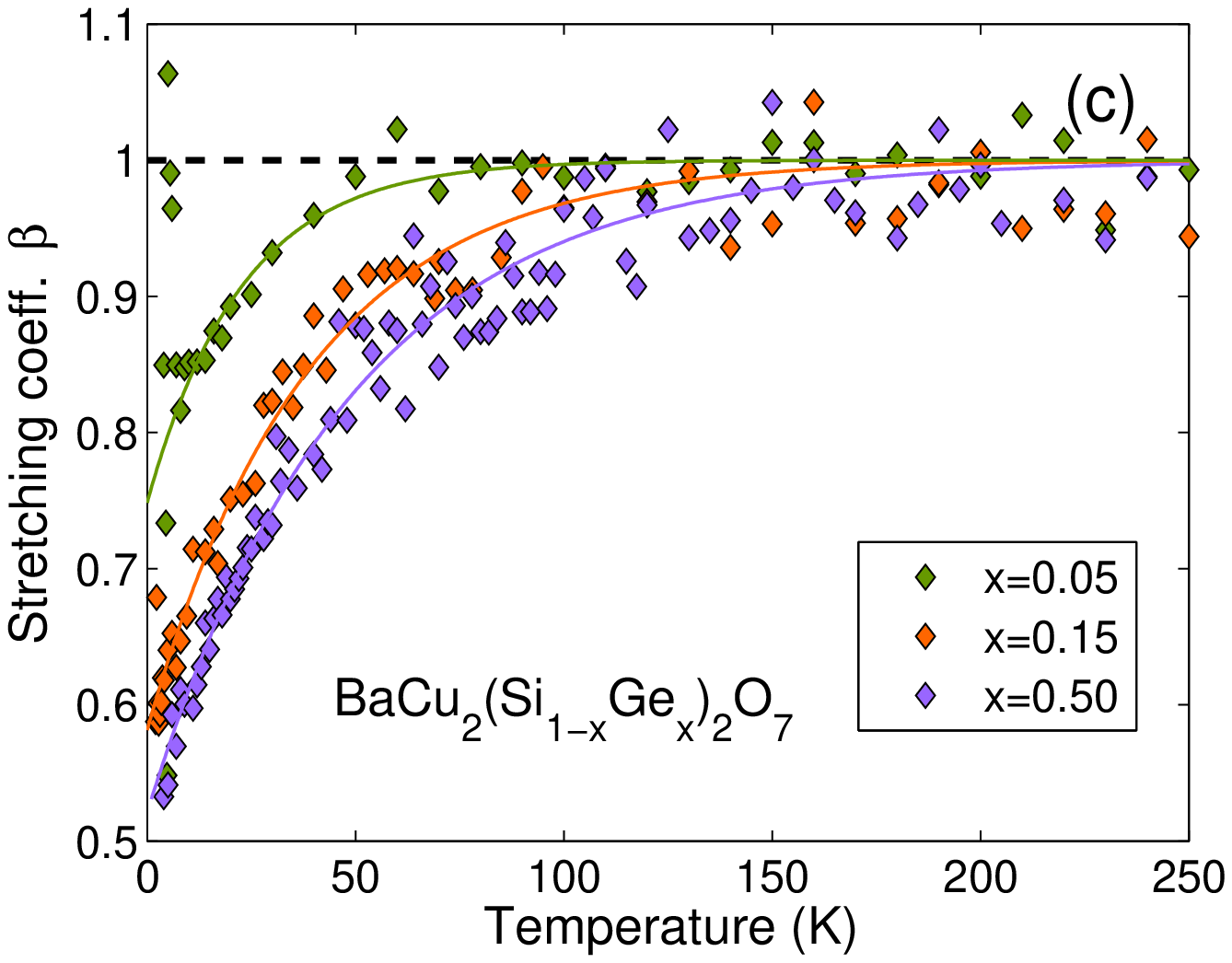} 
\caption{(a) Raw relaxation data and relevant fits for an $x=0.15$, $b$-axis 
oriented \BASGX\ single crystal. The mul\-ti\-de\-ca\-de recovery of 
magnetization at low temperatures reflects the significant decrease of 
the stretching coefficient $\beta$.
(b) Inverse spin-lattice relaxation time vs.\ temperature for a selection 
of \BASGX\ samples with different Ge-substitution values $x$. Inset: 
relaxation rates at low temperature. Peaks denote magnetic-transition 
values $T_\mathrm{N}$, arising from 3D interchain interactions, which 
decrease with increasing $x$.
(c) Stretching coefficient $\beta$ vs.\ temperature. Smooth lines are 
phenomenological fits. The horizontal dashed line indicates the $\beta = 1$ 
value expected in case of lack of disorder. Low-$T$ anomalies reflect the 
influence of the magnetic order with onset at $T_\mathrm{N}$.
\label{Fig:relax_15}
}
\end{figure}

\paragraph{NMR relaxation rates.}
In spin-chain compounds, the NMR spin-lattice relaxation rate $1/T_1$ has 
been shown to capture efficiently the subtle changes in spin fluctuations as 
a function of temperature.\cite{Shiroka2011} Below we demonstrate that 
this applies also to the case of a varying degree of disorder.

The nuclear spin relaxation is mainly driven by the electronic spin fluctuations 
on Cu$^{2+}$ ions, coupled to $^{29}$Si nuclei via the $q$-dependent 
hyperfine interactions $A(\boldsymbol{q})$: 
\begin{align}
& \frac{1}{T_{1z}} \propto \hbar^2 k_{\mathrm{B}} T  \gamma^2_n
\sum_{\substack{\alpha=x,y,z \\ \boldsymbol{q}}}
\left[ \tilde{A}^2_{x \alpha}(\boldsymbol{q}) + \tilde{A}^2_{y \alpha}(\boldsymbol{q}) \right] 
\frac{\chi''^c_{\alpha \alpha} (\boldsymbol{q},\omega_n)}{\omega_n} \label{eq:T1_relax}
\end{align}
Here $\omega_{n}$ is the resonance frequency, the $q$-dependent 
factor $A(\boldsymbol{q})$ is the NMR geometric form factor, while 
the $\omega$-dependent term 
$T \chi''^c_{\alpha \alpha} (\boldsymbol{q},\omega_n)/\omega_n  \propto S_{\perp}(\boldsymbol{q},\omega)$ 
is the dynamic structure factor.
In general, the nuclear relaxation rate involves both isotropic and 
anisotropic components of either hyperfine or dipolar origin.\cite{Azevedo1980}
As we discuss in Sec.~\ref{sec:discussion}, the latter, which mediate both 
transverse and longitudinal spin fluctuations, are not relevant in our case 
and do not appear in Eq.~(\ref{eq:T1_relax}), the standard expression. 
Thus, the longitudinal relaxation depends only on the transverse component 
of the fluctuating fields.\cite{Moriya1956}

Figure~\ref{Fig:relax_15}a shows the magnetization recovery curves 
$M(t)$ at different temperatures for the $x = 0.15$ case, with the 
crystalline $b$ axis oriented along the field. At a first approximation, 
the curves were fitted using $M(t) = 1 -\exp(-t/T_1)^\beta$, with $\beta$ 
a stretching exponent varying between 0 and 1. Clearly a lowering of temperature 
induces not only a shift of the curves towards longer delays but, most 
importantly, it implies an enhanced stretching (i.e., a reduction of $\beta$). 

An overview of the $1/T_1$ relaxation rates for different $x$ values is 
shown in Fig.~\ref{Fig:relax_15}b, the most important result in this work. 
Apart from the low-temperature ordered-phase region, the spin-lattice 
relaxation rate increases with temperature for all $x$ values. Its exact 
behavior depends on the degree of disorder, however. Thus, in the disorder 
free $x = 0$ case, $1/T_1(T)$ follows roughly a quadratic $T$ dependence. 
However, in the presence of disorder ($x > 0$), $1/T_1(T)$ increasingly 
tends to a linear-in-$T$ variation, which becomes close to perfect in the 
maximum-disorder $x =0.5$ case. 
In spite of recent attempts,\cite{Herbrych2013,Shu2018} the currently 
available theoretical models cannot successfully reproduce the experimental 
relaxation data for $x>0$, i.e., taking into account the random-singlet state. 
As for the magnetically ordered phase, the low-temperature results for 
various $x$ values are shown in the inset of Fig.~\ref{Fig:relax_15}b  
($x = 0.5$ is not included, since no ordering is observed above 1.5\,K). 
The sharp peaks observed at the relevant $T_\mathrm{N}$ values, 
reflect the critical slowing down of fluctuations near a second-order phase 
transition,\cite{Moriya1962,Borsa2007} which signal the antiferromagnetic 
transition. Clearly, the decrease of $T_\mathrm{N}$ with $x$, evident in 
the $1/T_1(T)$ data, fully confirms the magnetization results (see inset 
in Fig.~\ref{Fig:chi_T_x}).

The consequences of the random-singlet (RS) state are most clearly visible 
in Fig.~\ref{Fig:relax_15}c, where we report the evolution of the stretching 
exponent $\beta$ with temperature for different $x$ values. While for 
$x=0$ the magnetization recovery curves are always well reproduced by 
single-exponential ($\beta = 1$) fits (see Fig.~3b in Ref.~\onlinecite{Shiroka2011}), 
for $x=0.05$, $\beta$ is approximately 1 only down to $\sim 70$\,K. 
As $x$ increases, $\beta$ starts to depart distinctly from 1 at increasingly 
higher temperatures. This is confirmed by phenomenological 
$\beta(T) = [1-a\exp(-T/T_\mathrm{RS})]$ fits, which indicate higher 
$T_\mathrm{RS}$ values, i.e., the persistence of the RS regime over a 
wider $T$ range, as $x$ increases.
In previous work, we have shown that a stretched exponential relaxation, 
corresponds to a distribution of relaxation rates arising from a multitude 
of different local environments.\cite{Shiroka2011,Casola2012} In our case, 
this corresponds to coexisting electron-spin singlets with different coupling 
strengths, i.e., to the RS scenario. The earlier departure of $\beta$ from 
1 at higher $x$ values confirms that the RS regime emerges more easily 
in such cases.

\vspace{-5pt}\section{\label{sec:discussion}Discussion}
The NMR linewidths vs.\ temperature reported in Fig.~\ref{Fig:lines_15}, 
exhibit a very broad frequency/field distribution for all the $x >0$ cases, 
remarkably different from the pure $x = 0$ case, where the linewidths are 
almost constant with temperature.\cite{Casola2012} If the broadening 
were due to disorder effects only, at low $x$ values the $^{29}$Si nuclei 
would still probe (on average) practically the same environment as in the 
disorder-free case. However, the unusually large linewidths we observe 
already for $x = 0.05$, rules out the hypothesis of a disorder-induced 
low-temperature broadening. Instead, they strongly suggest that it is the 
RS pairing in the individual Cu$^{2+}$ chains to cause an extremely varied 
magnetic-field distribution (echoing the distribution of the inverse exchange 
constants), which is then reflected in the NMR linewidths. 

To understand the observed NMR relaxation rates, we recall that the 
typical 1D spin-chain $(\omega,\boldsymbol{q})$-dispersion, measured 
in inelastic neutron scattering, covers an area limited by a single sine 
curve from above and a double sine from below, the so-called two-spinon 
continuum.\cite{Arai1996,Zheludev2002,Zaliznyak2004,Lake2005} Due to the low radiofrequency 
energies employed in NMR (some $\mu$eV), the latter explores only the 
low-energy 1D spin excitations. Hence only the $q \approx 0$ and $q \approx \pi$ 
parts of the dynamical susceptibility $\chi''(T)$ contribute to the NMR relaxation. 
%
%
In the Luttinger liquid (LL) theory, $\chi''(T, q \approx 0)$ is 
temperature-independent\cite{Sachdev1994} implying that 
$1/T_1(T) \sim T$,\cite{Chitra1997} a result practically unaffected by 
form factor enhancements. At non zero temperatures, non-linear terms 
beyond the LL model imply a more complex behaviour, 
$1/T_1(T) \sim T \sqrt{T\ln^2(J/k_BT)/\omega_n}$.\cite{Sirker2009} 
However, since in our case the expected $1/\sqrt{H}$ dependence 
of $1/T_1$ was not observed experimentally, a model explaining also 
our observations is still missing.
As for $\chi''(T, q \approx \pi)$, this is divergent at zero-energies for 
$T \rightarrow 0$, as expected for a system with quasi-long-range 
AFM order, and its $\boldsymbol{q}$-integrated expression without 
the form factor would give $1/T_1(T) \sim T^{(1/2\alpha)-1}$, where 
$\alpha=1/2$ in zero field.\cite{Chitra1997} Since at low temperature 
the spectral weight is concentrated around the AFM point of the Brillouin 
zone, for $q \approx \pi$ one has to include also the NMR form factor, 
which finally gives a power-law dependence for $1/T_1(T)$ (see $x=0$ 
data in Fig.~\ref{Fig:relax_15}b). 

While the above considerations are strictly valid only for $x=0$, intermediate 
$x >0$ cases are characterized by an increasing  loss of translational 
invariance due to disorder. The translational inequivalence of sites introduces 
a multitude of \emph{local} nuclear relaxation times, notably different from 
the case of a chain with an average 
$J_{\mathrm{eff}} = (J_{\mathrm{Si}} + J_{\mathrm{Ge}})/2 \approx 37$\,meV 
value. This makes it difficult to account theoretically for the observed 
almost-linear $1/T_{1}(T)$ behavior. Indeed, the reported data are not 
described properly by existing theories,\cite{Herbrych2013,Shu2018} 
which even predict a reduction of $1/T_{1}(T)$ with increasing temperature.
%

We recall that there are two key problems in comparing experimental data 
with theoretical predictions: firstly, theory refers typically to nuclei belonging 
to the same ion (Cu$^{2+}$ in this case) as the electronic spins, while our 
experiment uses $^{29}$Si as a probe (the copper NMR signal relaxes too 
fast). Secondly, Ge sites have to be removed from the average and the effects 
of a nonzero magnetic field on low-$J$ bonds (increasingly important at low 
temperatures) have to be considered.
In addition, the obtained $T_1$ values contain not only the transverse 
spin fluctuations (which can be calculated), but also a contribution from 
anisotropic dipolar couplings $A_\mathrm{dip}(q)$, which mediate 
transverse and longitudinal spin fluctuations.\cite{Azevedo1980,Moriya1956} 
However, this is negligible in our field range, or can be minimized by a 
suitable orientation of the single crystal.\cite{Kuehne2009}
The above complexities make it clear why a reliable theory of $1/T_{1}(T)$ 
relaxation for 1D spin-chains with disorder is still not available. 
Nevertheless, by considering the experimental data and the translational 
invariance breaking for $x>0$, we can envisage a shift in spectral weight 
as the main cause for the disorder-induced modification of $1/T_{1}(T)$. 

The $x=0$ compound can sustain correlated antiferromagnetic fluctuations, 
whose main weight is at  $q \approx \pi$. As shown in Ref.~\onlinecite{Shiroka2011}, 
the experimental $^{29}$Si $1/T_1(T)$ data can only be fitted by including 
the NMR form factor which results in a power-law $1/T_1 \sim T^{\alpha}$ 
with $\alpha > 1$. A convincing presentation of the differing influences of 
$q = 0$ and $q = \pi$ excitations and of the site of the probing nuclei is 
given, e.g., in Ref.~\onlinecite{Thurber2001}. 
This representation, however, becomes ``blurred'' at $x >0$, 
implying a weight transfer towards $q \approx 0$, in turn reflected in a 
more linear $1/T_1$ behavior, as indeed observed.

%

\begin{figure}[t]
\includegraphics[width=0.85\columnwidth]{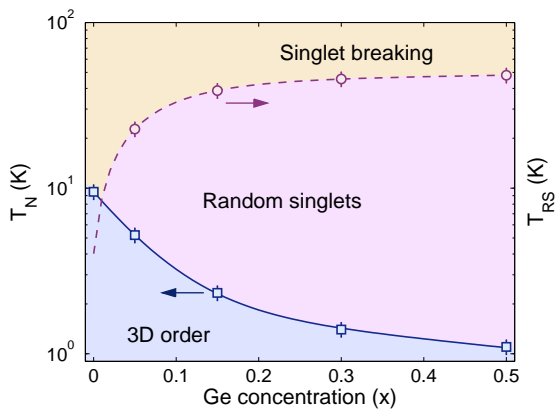} 
\caption{\label{Fig:phase_diag}Ordering temperature $T_\mathrm{N}$ and 
RS-regime crossover temperature $T_{\mathrm{RS}}$ (see text) vs.\ Ge concentration 
$x$. $T_{\mathrm{RS}}$ reaches its asymptotic value already for $x>0.1$, 
suggesting that even a weak degree of disorder is sufficient to achieve 
the random-singlet state. Schematically, the RS-regime exists between the 
lower 3D magnetic order phase boundary and the upper thermal singlet-breaking 
region. $T_\mathrm{N}$ data for $x = 0.5$ is from Ref.~\onlinecite{Thede2014}. 
Lines are guides to the eye, with the dashed line representing a crossover.}
\end{figure}

Since there is no agreed upon measure of randomness, we choose to 
represent our key results in Fig.~\ref{Fig:phase_diag} as a function of the 
Ge concentration $x$. Besides the ordering temperature $T_\mathrm{N}$, 
derived from the peak in both the NMR shift and the relaxation data, we 
report here also the crossover temperature $T_\mathrm{RS}$. The latter corresponds to the 
phenomenological temperature parameter appearing in 
$\beta(T) = 1 - a\exp(-T/T_\mathrm{RS})$, used to describe the stretching 
exponential data shown in Fig~\ref{Fig:relax_15}c. 
In view of its exponential $T$-dependence, $\beta(3T_\mathrm{RS}) = 0.97 \approx 1$. 
Hence the influence of quantum effects due to randomness may be regarded as negligible at this temperature.
The fast increase of $T_\mathrm{RS}$ 
from $x=0$ to 0.05, followed by a saturation above $x=0.15$, clearly 
indicates that the RS-regime sets in already at a small degree of disorder 
and it then persists  up to higher $x$ values. Such behavior is confirmed 
also by the raw relaxation-rate data in Fig~\ref{Fig:relax_15}b, where 
all the $1/T_1(T)$ curves for $x>0$ show a qualitatively similar behavior, 
distinctly different from $x=0$. The above results can be understood by 
considering that the average length of the chain segments with constant 
coupling $J_{\mathrm{Si}}$ becomes shorter as $x$ increases. 
At a macroscopic scale, also the magnetization curves show how a 
relatively small increase of disorder can produce significant changes 
in the physical properties. This is particularly evident in the low temperature 
regime, where $\chi(T)$ of the $x=0$ sample tends to a finite value, 
whereas those of the other compounds tend to diverge.

Finally, we recall that earlier muon-spin rotation ($\mu$SR) and specific-heat 
measurements, focusing on features of the ordered magnetic phase of \BASGX, 
came to similar conclusions.\cite{Thede2014} These studies found that, 
even in case of weak disorder, the magnetically-ordered state established 
below $T_\mathrm{N}$ is rather inhomogeneous. In addition, since in the 
presence of disorder the concept of qua\-si\-mo\-men\-tum and the 
related spin-wave excitation picture break down, the low-temperature 
($T \ll T_\mathrm{N}$) $x$-dependent specific-heat data, $C(T) \propto T^{\alpha}$, 
exhibit a strong decrease of $\alpha$ with increasing $x$ (at fixed temperature). 
The latter is a clear indication of an increasing density of low-energy 
excitations with $x$. These findings are compatible with our magnetization 
and NMR results, indicating \emph{for all $x>0$} the formation of an RS 
state above the respective ordering temperatures $T_\mathrm{N}$. 

\vspace{-5pt}
\section{Summary\label{sec:conclusion}}
By using magnetometry and nuclear magnetic resonance measurements 
we investigated the random Heisenberg spin-chain system \BASGX. 
Since a substitution of Si for Ge modulates the exchange couplings 
$J$ between Cu$^{2+}$ ions, by systematically varying the Ge content 
$x$, we could explore the transition from the purely ordered $x=0$ to 
the fully random $x=0.5$ case. 

The random-singlet (RS) regime, already demonstrated to occur at maximum 
disorder in the \BASGE\ case,\cite{Shiroka2011} was shown to emerge   
also for \emph{intermediate} $x$ values. Thus, for $x >0$, magnetic 
susceptibility curves show a low-temperature divergence, perfectly 
fitted by a random-singlet model with increasingly higher cutoff energy.
Similarly, both $^{29}$Si NMR relaxation rates $1/T_{1}(T)$ and the 
stretching coefficients $\beta(T)$ show clear indications of an RS 
scenario. These include a change in functional form and a departure 
from 1, respectively, both of which are compatible with the development 
of a multitude of spin-singlets with different coupling strengths. In all 
cases, the RS regime was shown to occur slightly above a (possible) 
3D magnetic-ordering temperature and to cover increasingly wider 
temperature ranges as $x$ increases (up to ca.\ 120\,K for $x=0.5$). 
Surprisingly, since the RS state sets in already in the $x=0.05$ case, 
we conclude that even a small degree of disorder is sufficient to induce 
it, albeit within a rather restricted $T$ range.

The here reported experimental evidence of the occurrence of an RS 
regime in spin-chains with a varying degree of disorder, is still missing 
a proper theoretical framework. Since to date only the fully random 
$x = 0.5$ case has been addressed, future theoretical efforts should 
consider also the more subtle cases of intermediate disorder.

\vspace{-5pt}\section*{Acknowledgments}\vspace{-5pt}
The authors thank A.\ Feiguin (Northeastern University, Boston) and 
Ch.\ R\"uegg (PSI) for useful discussions and A.\ Zheludev (ETH) 
for remarks on the first draft. The \BASGE\ samples used in this work 
were prepared in the early 2000's in the group of Prof.\ K.\ Uchinokura 
at the University of Tokyo. This work was supported in part by the 
Schweizerische Nationalfonds zur F\"{o}rderung der Wissenschaftlichen 
Forschung (SNF) through grant no.\ 200021-169455.


%

\end{document}